\documentclass[pra,english,preprintnumbers,amsmath,amssymb,nofootinbib,twocolumn,superscriptaddress]{revtex4-1}
\usepackage{braket}
\usepackage{multirow}

\usepackage[utf8]{inputenc}
\usepackage[T1]{fontenc}
\usepackage{amsmath}
\usepackage{amssymb}
\usepackage{graphicx}
\usepackage{tabularx}
\usepackage{bm}
\usepackage{pifont}
\usepackage{xcolor}

\def\0#1#2{\frac{#1}{#2}}

\def\s0#1#2{\mbox{\small{$ \frac{#1}{#2} $}}}

\newcommand{\beq}{\begin{equation}}
\newcommand{\eeq}{\end{equation}}
\newcommand{\bea}{\begin{eqnarray}}
\newcommand{\eea}{\end{eqnarray}}

\renewcommand{\vec}[1]{\mathbf{#1}}

\begin{document}

\title{Fermi-Fermi crossover in the ground state of 1D few-body systems \\ with anomalous three-body interactions}

\author{J. R. McKenney}
\affiliation{Department of Physics and Astronomy, University of North Carolina. Chapel Hill, North Carolina 27599-3255, USA}
\author{J. E. Drut}
\affiliation{Department of Physics and Astronomy, University of North Carolina. Chapel Hill, North Carolina 27599-3255, USA}

\begin{abstract}
In one spatial dimension, quantum systems with an attractive three-body contact interaction exhibit a scale anomaly. In this work, we 
examine the few-body sector for up to six particles. We study those systems with a self-consistent, non-perturbative, iterative method,
in the subspace of zero total momentum. Exploiting the structure of the contact interaction, the method reduces the complexity of 
obtaining the wavefunction by three powers of the dimension of the Hilbert space. We present results on the energy, and momentum and 
spatial structure, as well as Tan's contact. We find a Fermi-Fermi crossover interpolating between large, weakly bound trimers
and compact, deeply bound trimers: at weak coupling, the behavior is captured by degenerate perturbation theory; at strong coupling, 
the system is governed by an effective theory of heavy trimers (plus free particles in the case of asymmetric systems). Additionally, we 
find that there is no trimer-trimer attraction and therefore no six-body bound state.
\end{abstract}

\date{\today}

\maketitle 

\section{Introduction}
Quantum gases displaying scale invariance and scale anomalies have been under study both
theoretically and experimentally in a variety of contexts in the last two decades. 
In three dimensions (3D) the unitary Fermi gas is an example of a truly scale-invariant strongly interacting 
Fermi gas~\cite{ZwergerBook, Strinati:2018wdg} 
(in fact, it displays nonrelativistic conformal invariance~\cite{PhysRevD.76.086004}), whereas the two-dimensional (2D) 
version of the same system presents a scale anomaly~\cite{caruther, jackiwScaleIntro, JackiwDelta, holstein}. 
Both of the above examples have been realized experimentally with ultracold atoms by several 
groups and in particular the 2D case has been under intense scrutiny in recent years (see e.g.~\cite{PhysRevLett.105.030404,PhysRevLett.106.105301,PhysRevLett.108.235302,1367-2630-13-11-113032,PhysRevLett.108.070404,PhysRevLett.109.130403,PhysRevLett.112.045301,PhysRevLett.114.110403,PhysRevLett.114.230401,PhysRevLett.115.010401,PhysRevLett.116.045302}). The effect of the anomaly in 2D has also been extensively studied from the theoretical side 
(see e.g.~\cite{PhysRevA.55.R853, PhysRevLett.109.135301, PhysRevLett.106.110403, hoffman, PhysRevA.88.043636, PhysRevA.92.033603, 
PhysRevLett.115.115301, ORDONEZ201664, The2dPaper}, and~\cite{Rev2DLevinsenParish} for a review)

In the recent work of Ref.~\cite{PhysRevLett.120.243002}, it was shown that a very simple
generalization of the 2D scale-anomalous Fermi gas mentioned above can be achieved in 
one dimension (1D) as well. There, a three-species system of fermions, with an interaction fine-tuned such that only 
three-body forces are present, contains a dimensionless coupling constant that induces
a bound state (trimer) at arbitrarily small couplings. At about the same time, the bosonic analogue 
was studied by several groups in Refs.~\cite{PhysRevA.97.061603, PhysRevA.97.061604, PhysRevA.97.061605, PhysRevA.97.011602}, 
where the few- to many-body properties were
explored both analytically and numerically. In particular, such bosons were shown to form many-body 
bound states (``droplets''~\cite{PhysRevA.97.011602}) whose energy grows faster than exponentially 
with the particle number.

In contrast to the bosonic case mentioned above, relatively little is known about fermions. For instance, 
Ref.~\cite{PhysRevLett.120.243002} only addressed the fully symmetric three-body problem (i.e. the $1+1+1$ problem) 
and left open the questions of whether two fermionic trimers are attractive or repulsive (which would presumably leave an 
imprint on the nature of the many-body ground state), and whether there is a coupling-dependent crossover in the many-body behavior.

These types of questions are also relevant from a more general perspective, as three-body forces play 
a central role in nuclear physics. Indeed, two-body forces are typically regarded as the most important feature of interacting 
models, but three-body forces are being acknowledged as increasingly relevant~\cite{RevModPhys.85.197}.

In this work, we seek to address the properties of few-body, fermionic systems with three-body contact interactions in 1D.
Because the Pauli exclusion principle prohibits contact interactions between two fermions of the same flavor, the minimum nontrivial 
number of fermionic components for this model is three, which we employ. 
Although it is likely a challenge to engineer three-body forces, multiflavor experiments with SU($N$) symmetry have been underway for a 
few years~\cite{0034-4885-77-12-124401}, in particular for $N=3$~\cite{PhysRevLett.101.203202,ModawiLeggett,PhysRevLett.103.130404}. Here, we investigate 
specifically the energetics and structure of all possible interacting cases with up to six particles, labeling them by the numbers of fermions of 
each component, e.g., $2+2+2$ indicates two fermions of each flavor.

Conventionally, 1D few- and many-fermion problems can be solved exactly by way of the Bethe ansatz. The presence
of three-body interactions, however, makes this problem effectively 2D when particles approach the interaction region,
as we showed in Ref.~\cite{PhysRevLett.120.243002} (see also below). As a consequence, the problem addressed here requires a different kind of 
computational method. Fortunately, contact interactions allow for simplifications which, as we will show below, make 
few-body problems tractable up to six particles with modest computational resources. For those cases, the full wavefunction 
is in fact accessible.

Our approach bears conceptual similarities to the Faddeev-Yakubovsky equations~\cite{Faddeev}, where the wave function is separated into interacting and noninteracting pieces,
as well as to the Skorniakov-Ter-Martirosian equations~\cite{STM}, where a three-body problem is reduced to a lower-dimensional integral equation.
Our method is also reminiscent of the Lippmann-Schwinger equation~\cite{Sakurai}.

The remainder of the paper is organized as follows:
In Sec.~\ref{Sec:Hamiltonian}, we describe the Hamiltonian of the model, provide an example of the simplest case,
and explain how the coupling is renormalized.
Section~\ref{Sec:Formalism} introduces our computational method at length and provides details about the numerical implementation.
Results obtained are presented in Sec.~\ref{Sec:Results}, and we summarize and conclude in Sec.~\ref{Sec:Conclusions}.

\section{Hamiltonian and Renormalization \label{Sec:Hamiltonian}}
The system is defined by the following Hamiltonian
\beq
\hat H = \!\!\!\!\sum_{s=1,2,3}\int dp \; \epsilon(p) \; \hat a_{s,p}^\dagger \hat a_{s,p} + g \int dx \; \hat n_1(x)  \hat n_2(x)  \hat n_3(x),
\eeq
where $\epsilon(p) = \frac{\hbar^2p^2}{2m}$.
Here, $\hat a_{s,p}^\dagger$ and $\hat a_{s,p}$ are the fermionic creation and annihilation operators
for particles of species $s$ and momentum $p$, and $\hat n_s(x)$ is the corresponding density
at position $x$. Throughout this work, we will take $\hbar = k_B = m = 1$. 

In what follows, we use the momentum (position) variables $p, k, q$ (respectively $x, y, z$) to distinguish between fermionic species. For instance, in first quantization, the Hamiltonian
for the four-body $2+1+1$ system is
\beq
\label{Eq:H211}
\hat H = \frac{1}{2} \left(\hat p_1^2 + \hat p_2^2 + \hat k^2 +  \hat q^2\right) + g\sum_{i=1}^{2} \delta(x_i - y) \delta(y-z).
\eeq
For other numbers of particles, one interaction term must be included for every possible combination of three different-flavored fermions (trimers). 
Fermions of the same species do not interact because of the Pauli exclusion principle.

In order to renormalize the bare coupling $g$, we choose a lattice regularization, such that from this point on the bare coupling 
$g$ is a lattice coupling. The connection between $g$ and the specific physical situation is determined by a renormalization prescription,
which relates $g$ to the lattice momentum cutoff $\Lambda$ and the trimer binding energy $\epsilon_B$.
That relationship is obtained by solving the three-body problem as shown in Ref.~\cite{PhysRevLett.120.243002},
which yields
\beq
\label{Eq:glatt}
\frac{1}{g} = -\frac{1}{L^2} \sum_{\bf k} \frac{1}{\epsilon_{\bf k} + \epsilon_B},
\eeq
where $L = N_x \ell$ is the lattice size, $\ell$ is the lattice spacing, $\epsilon_{\bf k} = (k_1^2 + k_2^2 +k_3^2)/2$, 
${\bf k} = (2\pi/L)(n_1,n_2,n_3)$, and the sum covers $0 \leq |n_1 + n_2| \leq \Lambda$, with the constraint $n_1 + n_2 +n_3 = 0$ 
(i.e. vanishing total momentum). 

As can be appreciated in Eq.~(\ref{Eq:glatt}), the ultraviolet (i.e. short-range) behavior of the momentum sum is the same as that 
of the 2D two-body problem, in the sense that they both diverge logarithmically~\cite{PhysRevLett.120.243002}. We thus
see that, even though our problem is in 1D globally, it is in 2D locally.

\section{Formalism\label{Sec:Formalism}}

As mentioned above, in this work we consider systems with three species of fermions $1,2,3$ and corresponding varying particle numbers $N_1,N_2,N_3$.
We will fix those numbers and refer to the corresponding system as the $N_1+N_2+N_3$ problem. As species $1,2,3$ are distinguishable from
one another but particles are otherwise identical, we use as a starting point a product wavefunction with one factor for each species, i.e.
we start with a multiparticle wavefunction of the form $\psi = \psi_1 \psi_2 \psi_3$.

Within a given species, the wavefunction must be antisymmetric with respect to particle exchange, and to impose that property
we expand that factor in a basis of Slater determinants of plane waves. (In particular, we use a single Slater determinant as a starting point
in the iteration process described below.) Since the Hamiltonian preserves particle symmetry and the 
symmetric and antisymmetric subspaces are orthogonal, this initial projection 
is sufficient to capture the antisymmetry of the true wave function of the interacting system. In particular, when transforming 
from the position-space wave function $\psi(\vec{x})$ to the momentum-space wave function $\phi(\vec{p})$, the usual 
(unsymmetrized) Fourier transform may be used.
Additionally, we make use of the fact that the total momentum is conserved and impose a zero-total-momentum constraint on the
multiparticle wavefunction.

Formally, our approach consists in separating the kinetic and potential energy operators and grouping the former with
the energy eigenvalue $E$ in the Schr\"odinger equation, i.e. if $\hat H = \hat T + \hat V$, then we write the eigenvalue problem as
\beq
\phi = \frac{1}{\hat T - E}\hat V \phi
\eeq
where $\phi$ is the multiparticle wavefunction and the inverse of $\hat T - E$  is easily addressed as it is diagonal in momentum space.
The problem now becomes that of finding, for a proposed $E$, whether such a unit-eigenvalue equation can be satisfied. To that end,
we implement the iterative approach described below which, crucially, exploits the zero-range nature of the interaction to avoid computing determinants (which
would be the conventional way to solve the above problem).



As an illustration, consider the simplest non-trivial case of $2+1+1$ particles. The kinetic energy terms are simply given by the
sum of the kinetic energies of the individual particles. As shown in Eq.~\ref{Eq:H211}, the interacting part of the Schr\"odinger equation contains two 
terms corresponding to the two different ways in which the $1+1$ subsystem can interact with the $2$ identical particles.
We consider just one of those terms and proceed in momentum space.

One of the identical particles is a spectator, while the other three interact, such that the Fourier transform over the former is trivial. 
As usual, translation invariance yields a delta function that imposes conservation of total momentum of the reactants, leading to 
\beq
\frac{g}{L^2} \sum_{p'_1,k',q'} \phi(p'_1,p_2,k',q')\ \delta\left(P_1 - P_1' \right),
\eeq
where $P_1 = p_1+k+q$ and $P_1' = p'_1+k'+q'$.
While explicitly a function of all four momentum variables, this term can be reduced to a function of only one by choosing the center-of-momentum frame 
(i.e., only eigenstates of the total momentum operator with eigenvalue zero are considered), 
such that $P_1 + p_2 =0$; from here on out, we will work in this frame. Then, the delta function appears as $\delta\left(P'_1 + p_2\right)$.
Including both contributions to the interacting piece and rearranging, the full Schr\"odinger equation becomes
\beq\label{wf}
\phi(p_1,p_2,k,q) = G_0(P^2) \sum_{i} (-1)^{i} f(p_i),
\eeq
where the fermionic antisymmetry is manifest.
Here $P^2 = \frac{1}{2}\left(p_1^2 + p_2^2 + k^2 +q^2 \right)$ is the total kinetic energy, and
\beq 
G_0(P^2) = \frac{-g/L^2}{P^2 - E}
\eeq
is the noninteracting propagator multiplied by $-g/L^2$. The function $f$, not yet determined, is related to $\phi$ by
\beq\label{f_formula}
f(p) = \frac{1}{2!} \sum_{u,v} (1-\hat{P}_{up}) \phi(u,p,v,\varsigma),
\eeq
where $u,v$ are dummy variables, and the first sum has been carried out using the delta function;
the operator $\hat{P}_{up}$ exchanges $u$ and $p$.
Here and in what follows, $\varsigma$ denotes the negative of the sum of all other momentum variables present as arguments,
which results from momentum conservation; thus, in the above equation, $\varsigma = -u-p-v$.

The benefit of this approach is apparent already in Eq.~(\ref{wf}) upon considering the indistinguishability of fermions of the same species. 
Indeed, together with zero total momentum and the freedom to relabel summation indices, all interaction terms were
brought into the same functional form using the function $f(p)$ of Eq.~(\ref{f_formula}), which depends on only one momentum variable.

By substituting Eq.~\eqref{wf} into~\eqref{f_formula}, an implicit summation equation for $f$ is obtained which can be solved iteratively. 
Being a function of only one variable instead of four, $f$ is much more amenable to an iterative solution than the original equation in terms of the
momentum-space wavefunction $\phi$.
This three-dimensional reduction in the complexity of the problem that must be solved is a general feature of this method for the three-body contact interaction.

\begin{table*}
\caption{\label{tab1}
Wave functions and auxiliary functions for various systems. The collective index $(p_1 p_2 u)$, appearing in sums in the third column, indicates a summation over cyclic permutations of $p_1,\ p_2,\ u$ as the first three arguments of $\phi$, which correspond to like-flavor fermions. All $f$ expressions are summed over $u$ and $v$.
The argument of $G_0(P^2)$ is always the sum of the kinetic energies of all the particles in the system, i.e. $P^2 = \frac{1}{2}\left( \sum_{i} p_i^2 + \sum_{j} k_j^2 +\sum_{k} q_k^2 \right )$.
}
\begin{tabular}{c|c|c|c}
\hline
\hline
System & $\phi$ & $f$ summand & Normalization summand \\
\hline
2+1+1 & $\displaystyle G_0 \sum_{i} (-1)^{i} f(p_i)$ & $\displaystyle \frac{1}{2!} (1-\hat{P}_{up}) \phi(u,p,v,\varsigma)$ & $\displaystyle G_0^2 f(p_1) \sum_i(-1)^{i+1} f(p_i)$ \\
\hline
2+2+1 & $\displaystyle G_0\sum_{i,j} (-1)^{i+j} f(p_i,k_j)$ & $\displaystyle \frac{1}{(2!)^2} \left(1-\hat{P}_{up}\right)\left(1-\hat{P}_{vk}\right)\phi(u,p,v,k,\varsigma)$ & $\displaystyle G_0^2 f(p_1,k_1)\sum_{i,j} (-1)^{i+j} f(p_i,k_j)$ \\
\hline
3+1+1 & $\displaystyle G_0 \sum_{i,j,k} \epsilon_{ijk} f(p_i,p_j)$ & $\displaystyle \frac{1}{3!}\sum_{(p_1 p_2 u)} \phi(p_1,p_2,u,v,\varsigma)$ & $\displaystyle 
G_0^2 f(p_1,p_2) \sum_{i,j,k} \epsilon_{ijk} f(p_i,p_j)$ \\
\hline
2+2+2 & $\displaystyle G_0 \sum_{i,j,k} (-1)^{i+j+k} f(p_i,k_j,q_k)$ & \begin{tabular}{@{}c@{}} $\displaystyle \frac{1}{(2!)^3}\left(1-\hat{P}_{up_1}\right)
\left(1-\hat{P}_{vk_1}\right)\times$ \\ $\displaystyle \left(1-\hat{P}_{\varsigma q_1}\right)\phi(u,p_1,v,k_1,\varsigma,q_1) $\end{tabular}  & $\displaystyle G_0^2 f(p_1,k_1,q_1) \sum_{i,j,k} (-1)^{i+j+k+1} f(p_i,k_j,q_k)$ \\
\hline
3+2+1 & $\displaystyle G_0 \sum_{i,j,k,l} (-1)^l \epsilon_{ijk} f(p_i,p_j,k_l)$ & $\displaystyle \frac{1}{3!\cdot 2!} \sum_{(p_1 p_2 u)} \left(1-\hat{P}_{vk_1}\right)
\phi(u,p_1,p_2,v,k_1,\varsigma)$ & $\displaystyle G_0^2 f(p_1,p_2,k_1) \sum_{i,j,k,l} (-1)^{l+1} \epsilon_{ijk} f(p_i,p_j,k_l)$ \\
\hline
4+1+1 & $\displaystyle G_0 \sum_{i,j,k,l} \epsilon_{ijkl} f(p_j,p_k,p_l)$ & 
	\begin{tabular}{@{}c@{}} $\displaystyle \frac{1}{4!}\left\{1-\left[1-\left(1-\hat{P}_{up_3}\right)\hat{P}_{up_2}\right]\hat{P}_{up_1}\right\} \times $ \\ 
$\displaystyle \phi(u,p_1,p_2,p_3,v,\varsigma)$ \end{tabular}& 
$\displaystyle G_0^2 f(p_1,p_2,p_3) \sum_{i,j,k,l} \epsilon_{ijkl} f(p_i,p_j,p_k)$ \\
\hline
\hline
\end{tabular}
\end{table*}

Up to this point, the energy eigenvalues have not been addressed. By requiring the momentum-space wave function $\phi$ to be normalized, however, we obtain an equation that, when solved
simultaneously with the aforementioned summation equation for $f$, fixes $E$. A further simplification arises here:  Since all momenta are summed over, particle species are
interchangeable in $f$ inside the sums, greatly reducing the number of terms. Finally, $\phi$ (and $f$) may be taken to be real,
since any imaginary part would obey the same equation. All told, the normalization condition which fixes $E$ is
\beq\label{norm}
1 = \sum_{p_1,p_2,k,q} G_0^2(P^2) f(p_1) \sum_i(-1)^{i+1} f(p_i),
\eeq
where constant prefactors have been omitted for reasons that will be discussed in the next section.
With periodic boundary conditions, the kinetic contributions are expanded as $\left(\frac{2\pi n}{L}\right)^2$, $n$ integer, and the energy is expressed in terms of 
the three-body energy as $-E = \alpha \epsilon_B$, where $\epsilon_B$ is the (positive) binding energy of the trimer at the given coupling.

The different five- and six-body systems can be treated in a similar fashion; the corresponding expressions for all possible interacting configurations 
are collected in Table~\ref{tab1}. In all cases, when constructing the definition of $f$, it is crucial to do so in a way that retains the ordering of like-flavor fermions
as arguments of $\phi$.

\subsection{Iterative method}

For all cases, the wave function $\phi$ is initialized as a uniform momentum distribution subject to the constraints of antisymmetrization and zero total momentum.
The auxiliary functions $f$ are constructed from such $\phi$ (see Table~\ref{tab1}, third column) before being
fed into their defining implicit equations (see Table~\ref{tab1}, second and third columns). An initial run of 4000 iterations on $f$ is carried out before continuing until 
the lowest-energy value of $\alpha = -E/\epsilon_B$, which leads to a properly normalized wave function (built from $f$) is converged to a tolerance of $10^{-3}$. 
The numerical value of $\alpha$ is determined by cubic spline interpolation of the normalization of $\phi$ in a monotonic neighborhood containing unity.

To ensure numerical stability, $f$ is divided by the most recent normalization of $\phi$ before each iteration. 
Since the implicit summation equation is unaffected by overall factors (by virtue of the fact that the Schr\"odinger equation is linear) this 
division has no impact on the functional form; thus, constant prefactors, including those proportional to $g/L^2$ and those arising from combinatorics, 
are absorbed into
the overall scale of $f$ and thus may be omitted from the normalization expressions (all coupling dependence is carried and enforced by the implicit equation for $f$). 
While all values of $\alpha$ of course lead to proper normalizations immediately after this step, only the correct $\alpha$ is a fixed point and
remains equal to unity after repeated iterations. While the outermost momentum loops are carried out in parallel, all other entries of $f$ are updated as soon as 
they are computed, as in the Gauss-Seidel method.

\subsection{Weak-coupling expansion}

In the limit of very small coupling, we expect that the ground state can be expressed as a linear combination of noninteracting ground states. 
In this limit, the energy can be written as $E = E_{FG} - \epsilon$, where $E_{FG}$ is the energy of the noninteracting free gas (FG) ground state, and $\epsilon$ is a small deviation 
of the order of $\epsilon_B$.
At leading order in $\epsilon$, the surviving terms in the sum over $G_0(P^2)$ are the ones for which $P^2 = E_{FG}$. Similarly, at leading order in $\epsilon_B$,
$-g/L^2 \approx \epsilon_B$. As a result,
\beq
G_0(P^2) = \frac{\epsilon_B}{\epsilon} + \mathcal{O}\left(\frac{\epsilon_B}{P^2}\right).
\eeq
Working in the basis of noninteracting FG states, the momentum sums appearing in the equations for $f$ may now be carried out in closed form, allowing an exact solution at weak coupling.

\section{Results\label{Sec:Results}}

Using the technique described above, we have explored the properties of all attractively interacting systems with up to 6 particles beyond the already solved
case of 3 particles (namely the $1+1+1$ problem): 4 particles ($2+1+1$); 5 particles ($2+2+1$ and $3+1+1$); and 6 particles ($2+2+2$, $3+2+1$, and $4+1+1$).
In this section we show, for those systems, our results for the ground-state energy, spatial and momentum structure, and Tan's contact
of all six systems across five orders of magnitude in the trimer binding energy $\epsilon_B$ (expressed in dimensionless form as 
$\varepsilon_B \equiv \left(2\pi/L\right)^{-2} \epsilon_B$).

\subsection{Energy}
\begin{figure}[t]
\begin{center}
	\includegraphics[scale=0.30]{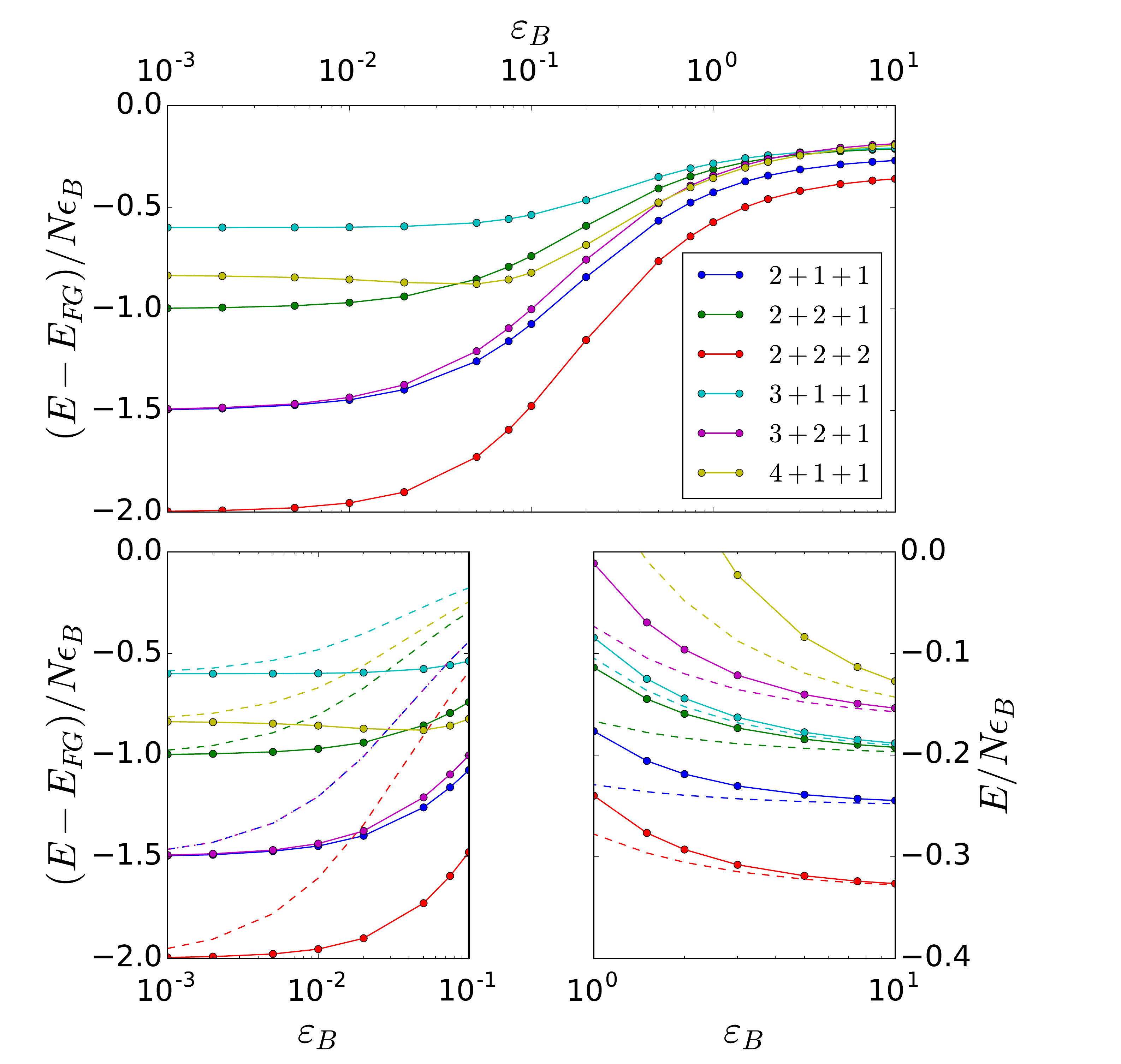}
	\caption{\label{Fig:energy} Top: Ground-state energy $E$ per particle of few-body systems as a function of the coupling, as measured by the binding energy $\varepsilon_B \equiv \left(2\pi/L\right)^{-2} \epsilon_B$ of the trimer. $E_{FG}$ is the ground-state energy of the corresponding noninteracting system. 
	Bottom left: Weak coupling regime, where dashed lines depict the NLO perturbative approximation using a linear combination of noninteracting
	ground states.
	Bottom right: Energy per particle at strong coupling in units of the trimer binding energy; the limiting values in this case are given by the number of trimers formed, divided by $N$, which yields (from bottom to top): $-1/3$, $-1/4$, $-1/5$, $-1/5$, $-1/6$, and $-1/6$.
	Dashed curves depict the NLO approximation in the particle-to-trimer mass ratio with Bethe-Peierls boundary conditions.}
\end{center}
\end{figure}

Figure~\ref{Fig:energy} shows our results for the ground-state energy $E$ of the various few-body problems as a function of the binding energy 
$\varepsilon_B \equiv \left(2\pi/L\right)^{-2} \epsilon_B$ of the trimer. $E_{FG}$ is the ground-state energy of the corresponding noninteracting system and 
$N=N_1 +N_2 + N_3$ is the total particle number. In all cases, the energy of the system is progressively more dominated by the binding energy of the trimer as 
the coupling is increased.

At weak coupling, one may attempt to capture the physics of the system with perturbation theory, which amounts
to computing $\langle \hat V \rangle$ in a noninteracting ground state, where $\hat V$ is the three-body interaction.
As we will show, however, the interaction term breaks the degeneracy among noninteracting ground states;
carrying out that calculation using a valid, but arbitrary, noninteracting ground state would not yield the correct result.
Instead (using a linear combination of noninteracting ground states) degenerate perturbation theory at next-to-leading order (NLO) yields the dashed lines 
of Fig.~\ref{Fig:energy} (bottom left), which agree with our non-perturbative results 
in the weak coupling limit. Our non-perturbative calculation converges to the following values (from bottom to top): $-2$ (red), $-3/2$ (blue, magenta), 
$-1$ (green), $-5/6$ (yellow), and $-3/5$ (cyan). The NLO energy of a single trimer would appear in all the panels of Fig.~\ref{Fig:energy} simply as $-1/3$, which is
actually (and deceptively) the exact result at all couplings.

To find the correct linear combinations of noninteracting ground states, we may solve the equations for $f$ at weak coupling as described previously.
For the $2+1+1$ system, the FG states only allow momentum values of $p=0$ and $|p| = 1$, allowing us to write
\beq 
\begin{pmatrix}
3 & -2 & -1 \\
-2 & 4 & -2 \\
-1 & -2 & 3
\end{pmatrix}
\begin{pmatrix}
f(-1) \\
f(0) \\
f(1)
\end{pmatrix}
= \frac{\epsilon}{\epsilon_B}
\begin{pmatrix}
f(-1) \\
f(0) \\
f(1)
\end{pmatrix},
\eeq
yielding eigenvalues of $\epsilon = 4 \epsilon_B$ and $\epsilon = 6 \epsilon_B$, where the total energy is $E = E_{FG} - \epsilon$; there is also a spurious solution of $\epsilon = 0$ that is inconsistent with the assumed approximation.
This splitting of energies demonstrates how an arbitrarily small coupling breaks the degeneracy of the FG states.
The eigenvector corresponding to the ground state, $\epsilon = 6 \epsilon_B$, demands that $f(-1) = f(1) = -\frac{1}{2} f(0)$. Consistency with the equations in Table~\ref{tab1}
then determines
\beq
f(p) = \sqrt{2}\left(\delta(p-1) - 2\delta(p) + \delta(p+1)\right),
\eeq
which also allows determination of the position-space wave function as
\beq
\psi(x_1,x_2,y,z) = \tilde{\psi}(x_2,y,z) - \tilde{\psi}(x_1,y,z),
\eeq
where $\tilde{\psi}(x,y,z) = \frac{1}{\sqrt{2} L^2}\left[\cos\left(\frac{2\pi (x-y)}{L}\right) + \cos\left(\frac{2\pi (x-z)}{L}\right)\right]$.
The same procedure may be carried out for the other systems; in all cases, the ground-state eigenvalues agree with the limiting values in Fig.~\ref{Fig:energy} (bottom left).

%

%
%
In the limit of strong coupling, trimers will be extremely localized and act as impenetrable, composite fermions. The low-energy effective Hamiltonian
in terms of noninteracting trimers and the remaining unbound fermions can be written as
\beq
\hat{H}_{\text{eff}} = \frac{1}{2m} \left( \hat{p}^2_{\text{unbound}} + \lambda \hat{p}^2_{\text{trimer}}\right),
\eeq
where $\lambda = 1/3$ is the fermion-to-trimer mass ratio and is treated as a small parameter. 
For all of the systems considered here except the $2+2+2$ problem, only one trimer forms. In those cases, the limit of small $\lambda$
yields a massive trimer which, using Bethe-Peierls boundary conditions, shows that the problem is equivalent to that of noninteracting fermions in a 1D hard-wall box.
Using that mapping, we present results on a NLO (first order in $\lambda$) approximation to the energy for comparison with our numerical 
results in Fig.~\ref{Fig:energy} (bottom right), for which we find the agreement to be excellent. The $2+2+2$ system comprises two trimers, 
so we treat its strong coupling limit as that of two identical fermions of mass $3m$ on a 1D ring; again the agreement is very satisfactory.
%

\subsection{Trimer-trimer interaction and (un-)bound state}

Of particular interest is the nature of the interaction between two trimers. At strong coupling, the deeply bound trimers likely repel each other due to Pauli exclusion of
their constituent fermions. At weak and intermediate couplings, however, it is \textit{a priori} possible for the trimer-trimer interaction to be attractive, such that hexamers may form,
and that trimer-trimer pairing may affect the nature of the ground state in the many-body regime.

\begin{figure}[t]
	\includegraphics[scale=0.3]{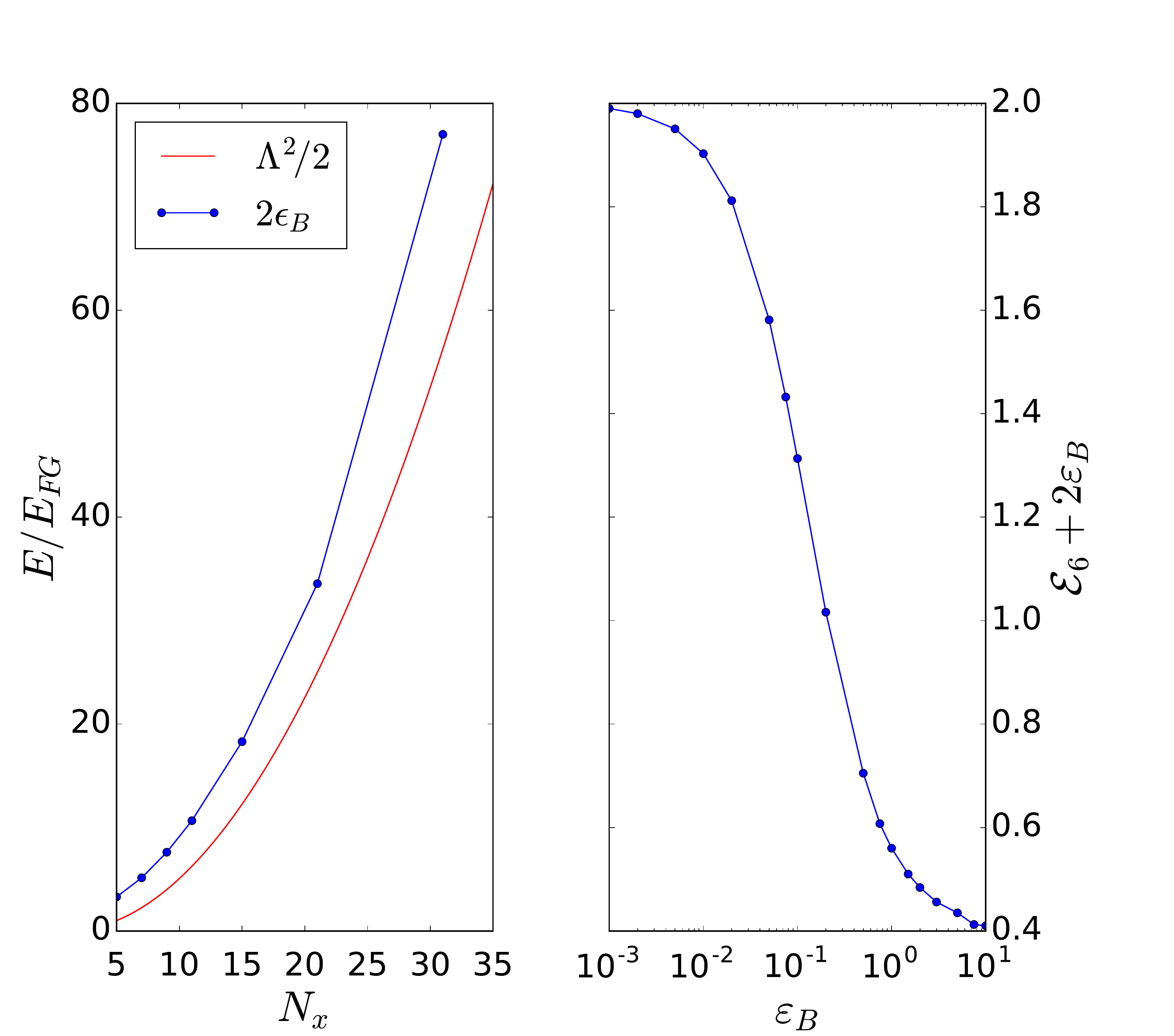}
	\caption{\label{Fig:hexamer} Left: Coupling ($\epsilon_B$) required for threshold trimer-trimer binding. In addition to diverging in the continuum limit, 
		the energy of the bound hexamer state (dotted blue) always exceeds the lattice cutoff energy (red), indicating that no such state exists.
		Right: Energy of $2+2+2$ system relative to the energy of two trimers, showing that the trimers repel each other.}
\end{figure}

To address this question, we apply our method to the $2+2+2$ system, fixing the energy to
twice the binding energy $\epsilon_B$ (the threshold for attractive interaction), and search for the smallest value of $\epsilon_B$ that allows a normalized wave function.
While this can be carried out numerically, our results indicate that such a hexamer state is not physical. In Fig.~\ref{Fig:hexamer}, the calculated energy of this state
is larger than the kinetic energy associated with the momentum cutoff $\Lambda$, violating the necessary separation of scales in our renomalization scheme.
Furthermore, in contrast to all other cases investigated, the energy does not converge to a continuum value in the limit of large $N_x$. Combined with the fact that the
$2+2+2$ system's energy appears to approach $2\epsilon_B$ asymptotically from below (see Table~\ref{tab2}), we conclude that the trimers approach a hard-core repulsive interaction
in the limit of large coupling.

Based on the above analysis, we conclude that at least at the few-body level there is no trimer-trimer bound-state formation. It remains to be determined whether
the presence of a Fermi surface (i.e. a finite background fermion density) induces trimer-trimer pairing in the many-body regime.

\subsection{Structure}

\subsubsection{Momentum structure}

\begin{figure}[t]
\begin{center}
	\includegraphics[scale=0.385]{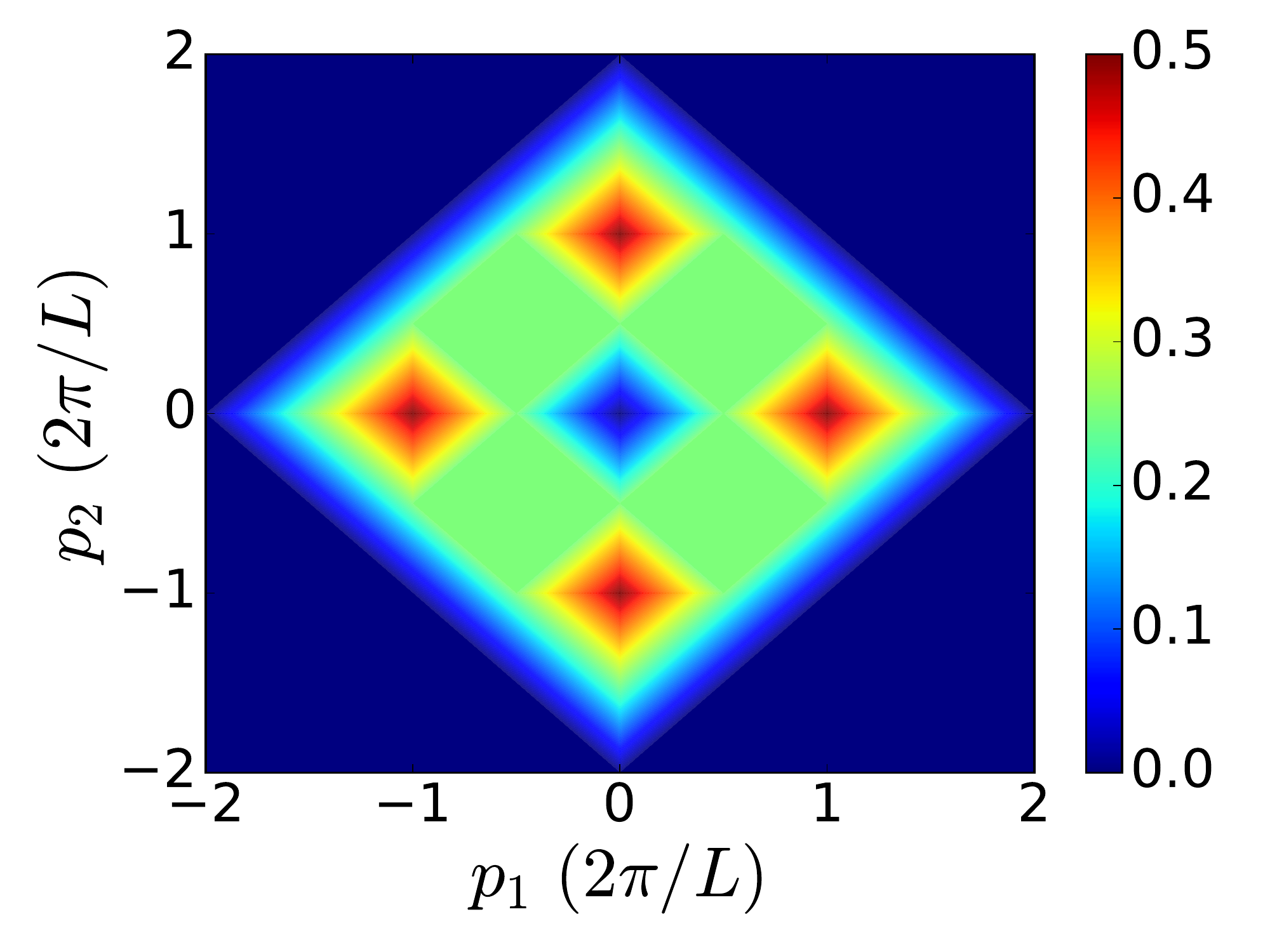}
	\caption{\label{Fig:weak_momentum} Weakly interacting $\left(\varepsilon_B = 10^{-4}\right)$ momentum density of two identical particles in the $2+1+1$ system with the distinct particles
	integrated out. Here, identical particles fill Fermi level rather than take on opposing momenta.}
\end{center}
\end{figure}

\begin{figure}[t]
\begin{center}
	\includegraphics[scale=0.385]{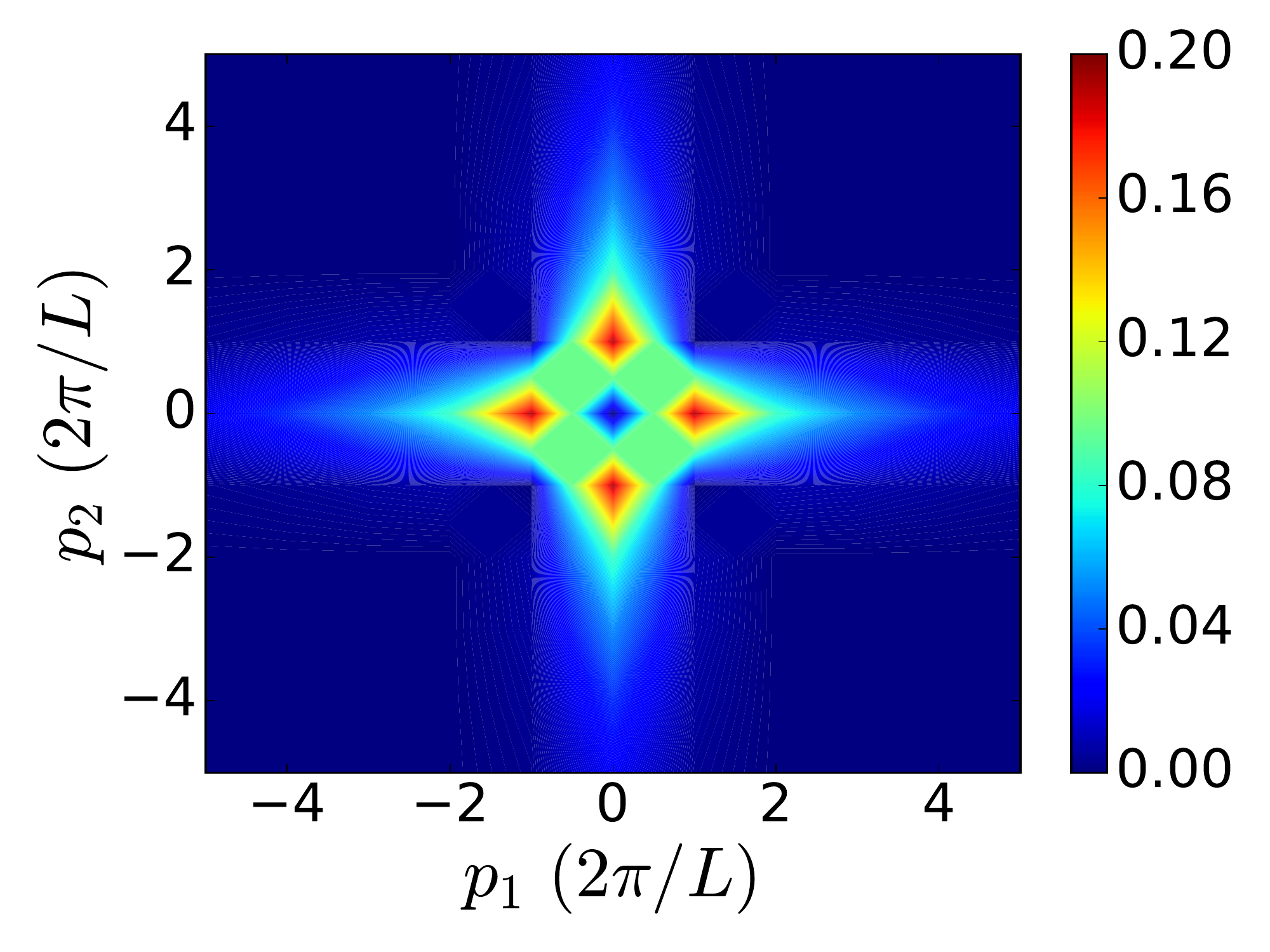}
	\caption{\label{Fig:strong_momentum} Strongly interacting $\left(\varepsilon_B = 10\right)$ momentum density of two identical particles in the $2+1+1$ system with the distinct particles
	integrated out.}
\end{center}
\end{figure}

In Figs.~\ref{Fig:weak_momentum} and~\ref{Fig:strong_momentum} we show the momentum distribution for the $2+1+1$ problem at
weak and strong coupling, respectively. As expected, at weak coupling, only low-lying momentum states are occupied. 
As we increase the coupling, higher momentum states become filled (Fig.~\ref{Fig:strong_momentum}; note the difference in scales compared to Fig.~\ref{Fig:weak_momentum}).
We anticipate that the increased number of momentum states contributing at strong coupling corresponds to enhanced spatial localization of trimers.
%

\subsubsection{Spatial structure: selected two-body density matrices}

\begin{figure}[t]
\begin{center}
	\includegraphics[scale=0.17]{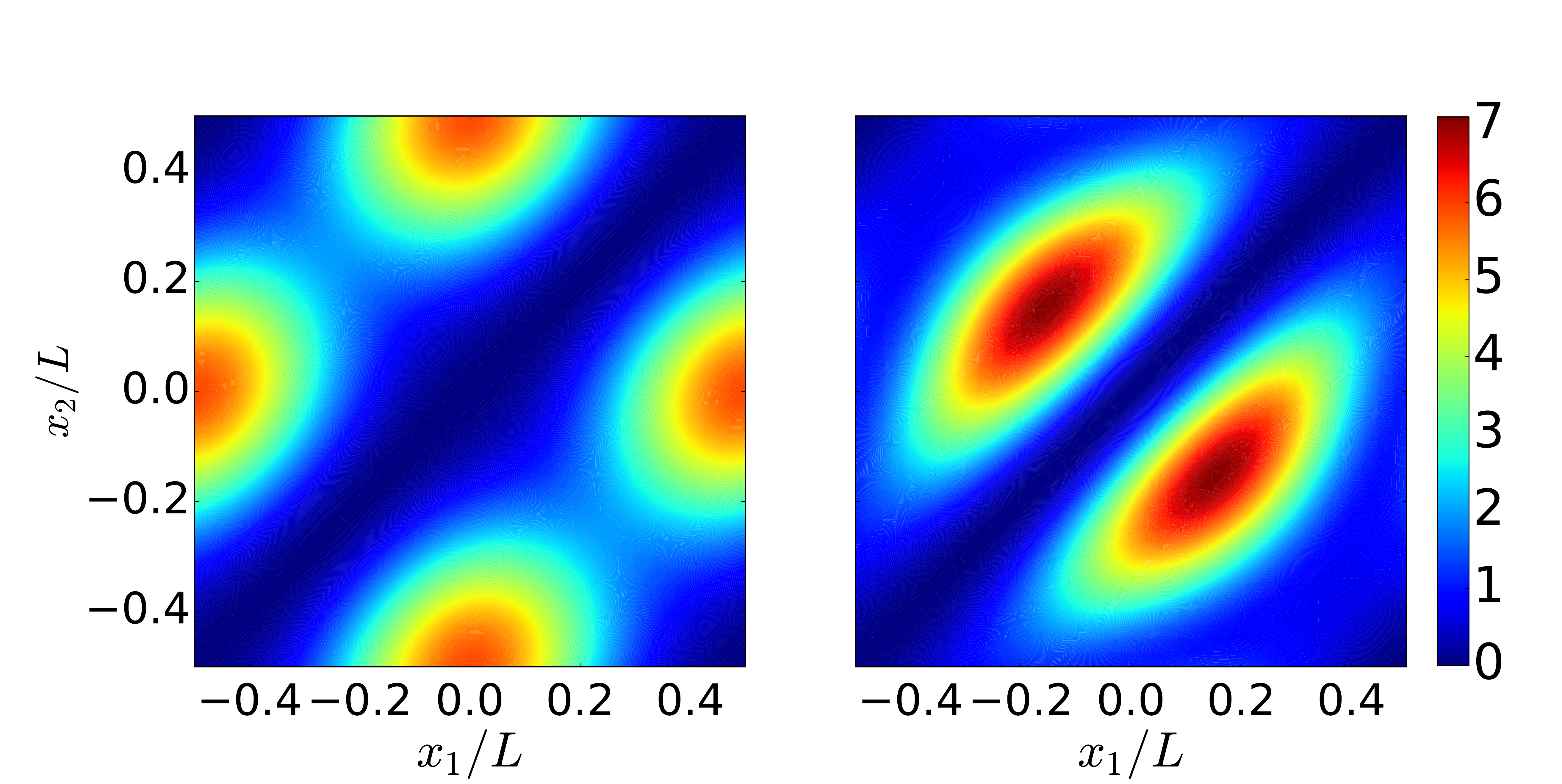}
	\caption{\label{Fig:gd_density} Weakly interacting particle density (left: $E - E_{FG}= -6\epsilon_B$; right: $E - E_{FG} = -4\epsilon_B$) of two identical particles in the $2+1+1$ system. 
	For the two distinct particles, one is fixed at the origin, and the other is integrated out.}
\end{center}
\end{figure}

\begin{figure}[t]
\begin{center}
	\includegraphics[scale=0.40]{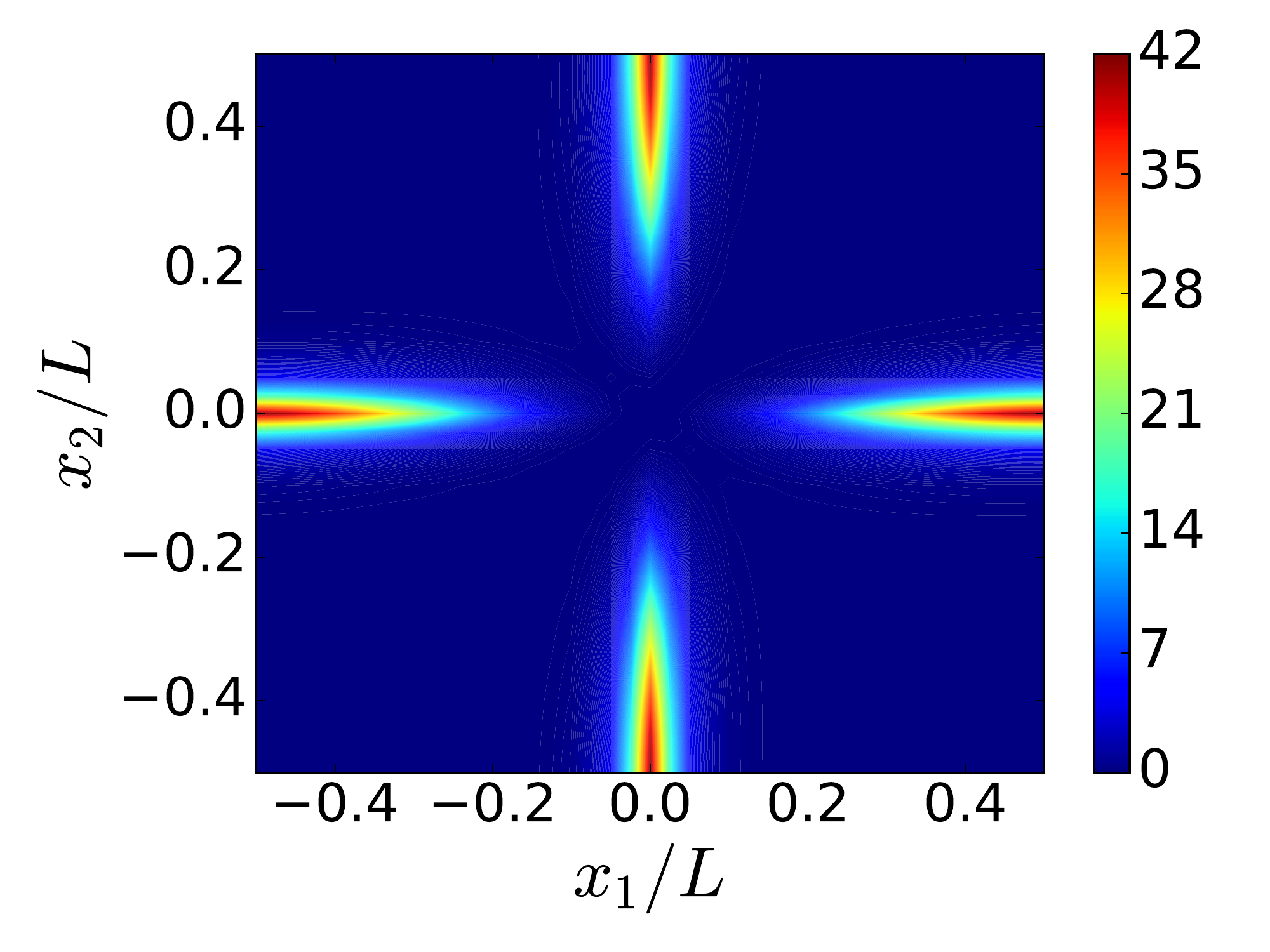}
	\caption{\label{Fig:strong_density} Strongly interacting $\left(\varepsilon_B = 10\right)$ particle density of two identical particles in the $2+1+1$ system. 
	For the two distinct particles, one is fixed at the origin, and the other is integrated out.}
\end{center}
\end{figure}

Since $f$ provides full information of the momentum-space wave function, we can also access the position-space wave function by Fourier transformation. Here, we provide spatial density distributions
of the simplest case ($2+1+1$) to illustrate how an excess particle positions itself around a trimer. In Figs.~\ref{Fig:gd_density} and \ref{Fig:strong_density}, we present the particle density 
of the two identical particles as a function of their positions. Since the system is translation invariant,
we fix the position of one of the unique fermions at the origin and integrate over the position of the other. With this choice, the (1D) origin serves as the center of the trimer, and
we will refer to it accordingly.

\begin{figure}[t]
\begin{center}
	\includegraphics[scale=0.36]{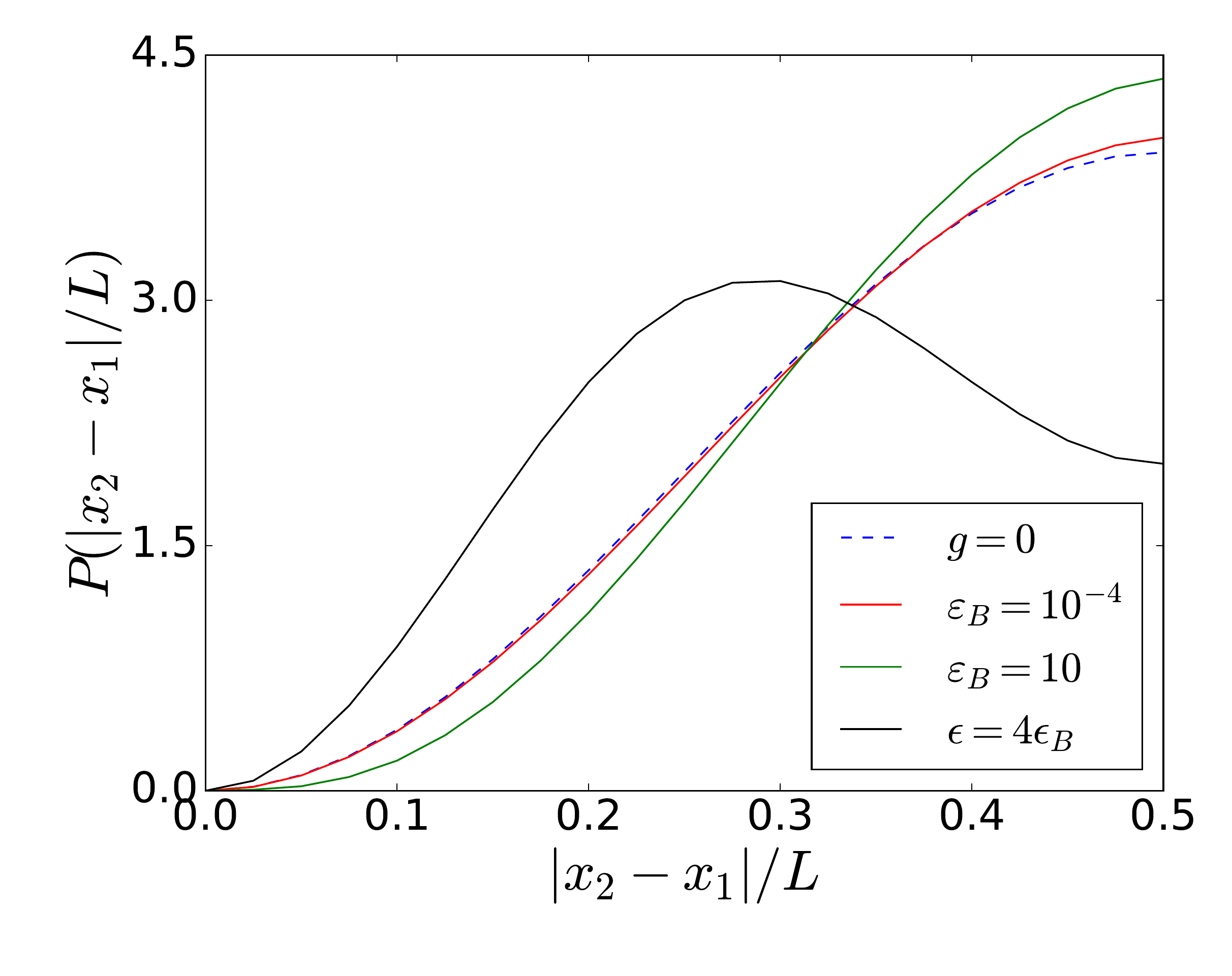}
	\caption{\label{Fig:separation} Probability $P(|x_2 - x_1|/L)$ of finding the two identical particles at a given separation $|x_2 - x_1|$ of the two identical particles in the 2+1+1 problem. This function is translation-invariant after integrating over all positions of the two distinct particles as in Eq.~(\ref{Eq:Separation211}). The stronger the interaction, the farther apart the particles are likely to be. }
\end{center}
\end{figure}

Figure~\ref{Fig:gd_density} shows the analytical results in the weakly interacting limit (our non-perturbative result at $\varepsilon_B = 10^{-4}$ is indistinguishable from the
perturbative result at $\epsilon = 6\epsilon_B$). The left panel reveals that the ground state favors a configuration of one large trimer far away from a free particle.
The first excited state, however, favors a four-body molecular configuration (right panel) where the two identical fermions are equidistant from the molecular center. 
The probability distribution 
\beq
\label{Eq:Separation211}
P(|x_2-x_1|/L) = \int dy\; dz\; |\psi(x_1,x_2,y,z)|^2,
\eeq
for particle separations $|x_2-x_1|$ is plotted in Fig.~\ref{Fig:separation}, showing that the ground state typically maximizes the distance between
identical fermions.

Figure~\ref{Fig:strong_density} provides support for our approximate model at strong coupling.
One fermion is tightly bound and highly localized around the center of the trimer, while the other tends to be on the opposite side of the ring (we use periodic boundary conditions), albeit with much more freedom to
move about the ring. In fact, due to the high degree of localization of the particle in the trimer, the separation of the two particles
can be approximately regarded as the position of the free fermion. The roughly sinusoidal shape of this curve in Fig.~\ref{Fig:separation} is thus 
yet another validation of our approximation at strong coupling.

%

\subsection{Contact}

\begin{figure}[h]
\begin{center}
	\includegraphics[scale=0.385]{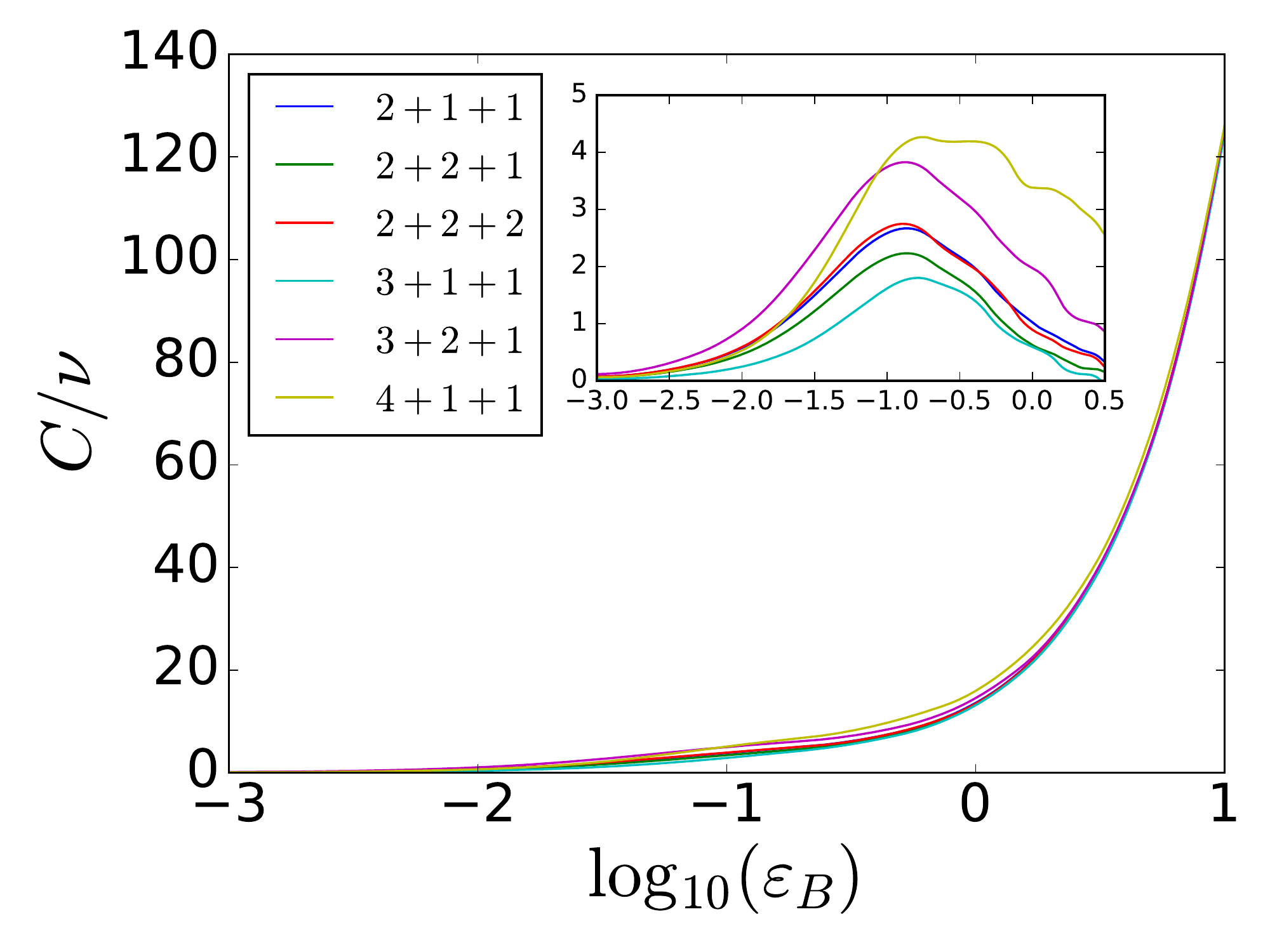}
	\caption{\label{Fig:contact} Tan's contact divided by the number of trimers present, $\nu$, as a function of the dimensionless trimer binding energy $\varepsilon_B$. Inset: Same as main plot, after subtracting the contact for $\nu$ trimers; the oscillations at strong coupling are due to interpolation effects
	on the energy upon numerical differentiation.}
\end{center}
\end{figure}

Tan's contact~\cite{TAN20082952, TAN20082971, TAN20082987} controls the high-momentum (short-distance) asymptotics of correlation functions in
theories with short-range interactions. While the resulting universal relations have not yet been derived for the present case, it is easy to see
using the Feynman-Hellman theorem that the contact can be expressed as the derivative of the energy with respect to the 
logarithm of the scattering length (as shown for the system considered here in Ref.~\cite{PhysRevLett.120.243002}). 
Here, we work in terms of the binding energy rather than the scattering length, such that the (dimensionless) contact takes the form
\beq
C = -4\pi \frac{\partial \mathcal{E}}{\partial\left(\ln \epsilon_B \right)},
\eeq
where $(2\pi/L)^2 \mathcal{E} = E$. We compute this derivative from a cubic spline interpolation of the energy and present the results in Fig.~\ref{Fig:contact}, 
where we divide by the number of whole trimers, $\nu$, that can be formed in each system. 
Though there are some small deviations (possibly stemming from the interpolation), the curves for all six cases overlap across the entire range of 
coupling values. The agreement is strongest at large coupling, indicating that the energy there is dominated by whole trimers---were there larger 
molecules forming, we would expect to see variation among the different systems. At intermediate coupling strengths, the modest differences may 
reflect competition between the different particles for inclusion in the trimer.

\section{Summary and Conclusions\label{Sec:Conclusions}}

To summarize, we have extended the analysis of the three-body problem with a three-body contact interaction of Ref.~\cite{PhysRevLett.120.243002} 
beyond the $1+1+1$ system. Here, we have studied the higher-body problems up to six particles in all possible non-trivial combinations, namely: 
4 particles ($2+1+1$); 5 particles ($2+2+1$ and $3+1+1$); and 6 particles ($2+2+2$, $3+2+1$, and $4+1+1$). With the exception of the $2+2+2$ case, which
will form two trimers, each of the cases considered will form only one trimer. 
To solve those problems, we designed a self-consistent numerical approach, which is the natural generalization of the analytic solution of the 
$1+1+1$ problem of Ref.~\cite{PhysRevLett.120.243002}.
Our solution gives immediate access to essential properties like the energy and Tan's contact, but it also provides the multiparticle 
wavefunction, from which everything else is calculable. 

As a function of the interaction strength, our results show that the energy of the various systems studied presents clear variations as
a function of the system composition (see Fig.~\ref{Fig:energy}). On the other hand, we have found that Tan's contact displays extremely small
variations across the different systems.
At weak coupling, we compare with first-order perturbation theory in the bare coupling, renormalized using the three-body binding energy,
and find good agreement after taking into account the breaking of the degeneracy of noninteracting ground states, i.e. using degenerate
perturbation theory.  At strong coupling, one expects a description dominated by bound trimers plus noninteracting particles.
We implemented that approximation by representing the trimer as a point-like fermion of mass $3m$ interacting with
the remaining particles via a hard-core potential, which we implemented simply as a vanishing boundary condition in the
many-body wavefunction (\`a la Bethe-Peierls). That description works remarkably well already at couplings of $\varepsilon_B = 10$,
as shown in Fig.~\ref{Fig:energy}.

Our calculations suggest that, as in the case of pure two-body contact forces, Pauli exclusion provides a strong repulsion that
overpowers any residual attraction coming from the three-body force. The trimers always experience a repulsive interaction, as far 
as we have explored. The emerging picture is therefore that of a crossover between the original, attractively interacting fermions 
(weakly bound into extended trimers) and deeply bound composite fermions (strongly localized trimers) with a residual repulsive interaction. 
Although we have not found trimer-trimer binding, it would be interesting to investigate the effect of a finite interaction range, which has been shown
to yield droplet formation and eventually lead to a liquid-gas transition in 1D systems with finite-range, two-body interactions 
(see e.g.~\cite{PhysRevLett.101.096401, 1712.03842}).


\begin{appendix}

\begin{table*}
\caption{\label{tab2} Numerical energy ratios $\alpha=-E/\epsilon_B$ corresponding to each coupling $\varepsilon_B = (2\pi/L)^{-2} \epsilon_B$.}
\begin{tabularx}{\textwidth}{X X X X X X X}
\hline
\hline
\multirow{2}{*}{$\varepsilon_B$} & \multicolumn{6}{c}{System} \\
\cline{2-7} 
& $2+1+1$ & $2+2+1$ & $3+1+1$ & $2+2+2$ & $3+2+1$ & $4+1+1$ \\
\hline
$10^{-3}$ & $-994.019$ & $-995.016$ & $-997.001$ & $-1988.024$ & $-1991.045$ & $-3994.985$ \\
$2\times 10^{-3}$ & $-494.038$ & $-495.031$ & $-497.002$ & $-988.048$ & $-991.085$ & $-1994.970$ \\
$5\times 10^{-3}$ & $-194.107$  & $-195.078$ & $-197.004$ & $-388.126$ & $-391.193$ & $-794.929$ \\
$10^{-2}$ & $-94.208$  & $-95.154$ & $-97.011$ & $-188.269$ & $-191.385$ & $-394.869$ \\
$2\times 10^{-2}$ & $-44.411$ & $-45.306$ & $-47.030$ & $-88.589$ & $-91.757$ & $-194.779$ \\
$5\times 10^{-2}$ & $-14.969$ & $-15.729$ & $-17.117$ & $-29.626$ & $-32.751$ & $-74.736$ \\
$7.5\times 10^{-2}$ & $-8.698$ & $-9.369$ & $-10.546$ & $-17.097$ & $-20.096$ & $-48.202$ \\
$10^{-1}$ & $-5.702$ & $-6.301$ & $-7.313$ & $-11.133$ & $-13.989$ & $-35.067$ \\
$2\times 10^{-1}$ & $-1.626$ & $-2.045$ & $-2.673$ & $-3.081$ & $-5.453$ & $-15.888$ \\
$5\times 10^{-1}$ & $0.264$ & $0.036$ & $-0.246$ & $0.590$ & $-1.116$ & $-5.144$ \\
$7.5\times 10^{-1}$ & $0.573$ & $0.403$ & $0.209$ & $1.190$ & $-0.306$ & $-2.925$ \\
$1$ & $0.706$ & $0.569$ & $0.422$ & $1.440$ & $0.068$ & $-1.864$ \\
$1.5$ & $0.823$  & $0.724$ & $0.625$ & $1.660$ & $0.417$ & $-0.837$ \\
$2$ & $0.875$ & $0.797$ & $0.721$ & $1.758$ & $0.577$ & $-0.341$ \\
$3$ & $0.922$ & $0.867$ & $0.815$ & $1.848$ & $0.729$ & $0.137$ \\
$5$ & $0.956$ & $0.922$ & $0.888$ & $1.913$ & $0.843$ & $0.503$ \\
$7.5$ & $0.972$ & $0.949$ & $0.924$ & $1.945$ & $0.896$ & $0.680$ \\
$10$ & $0.979$ & $0.962$ & $0.942$ & $1.959$ & $0.923$ & $0.764$ \\
\hline
\hline
\end{tabularx}
\end{table*}

\begin{table*}
\caption{\label{tab3} Lattice sizes $N_x$ used for each entry of Table~\ref{tab2}.}
\begin{tabularx}{\textwidth}{X X X X X X X}
\hline
\hline
\multirow{2}{*}{$\varepsilon_B$} & \multicolumn{6}{c}{System} \\
\cline{2-7} 
& $2+1+1$ & $2+2+1$ & $3+1+1$ & $2+2+2$ & $3+2+1$ & $4+1+1$ \\
\hline
$10^{-3}$ & $41$ & $15$ & $15$ & $15$ & $15$ & $15$ \\
$2\times 10^{-3}$ & $41$ & $15$ & $15$ & $15$ & $15$ & $15$ \\
$5\times 10^{-3}$ & $41$  & $15$ & $15$ & $15$ & $15$ & $15$ \\
$10^{-2}$ & $41$  & $41$ & $21$ & $15$ & $15$ & $15$ \\
$2\times 10^{-2}$ & $41$ & $41$ & $21$ & $15$ & $15$ & $15$ \\
$5\times 10^{-2}$ & $41$ & $41$ & $21$ & $15$ & $15$ & $15$ \\
$7.5\times 10^{-2}$ & $41$ & $41$ & $21$ & $15$ & $15$ & $15$ \\
$10^{-1}$ & $41$ & $41$ & $21$ & $15$ & $15$ & $15$ \\
$2\times 10^{-1}$ & $41$ & $41$ & $21$ & $15$ & $15$ & $15$ \\
$5\times 10^{-1}$ & $41$ & $41$ & $25$ & $15$ & $21$ & $21$ \\
$7.5\times 10^{-1}$ & $41$ & $41$ & $51$ & $31$ & $21$ & $31$ \\
$1$ & $41$ & $41$ & $51$ & $31$ & $21$ & $31$ \\
$1.5$ & $41$  & $41$ & $51$ & $31$ & $21$ & $31$ \\
$2$ & $41$ & $41$ & $51$ & $31$ & $25$ & $31$ \\
$3$ & $41$ & $41$ & $51$ & $31$ & $31$ & $31$ \\
$5$ & $41$ & $41$ & $61$ & $31$ & $31$ & $31$ \\
$7.5$ & $41$ & $41$ & $61$ & $41$ & $41$ & $41$ \\
$10$ & $41$ & $41$ & $61$ & $41$ & $41$ & $41$ \\
\hline
\hline
\end{tabularx}
\end{table*}

\end{appendix}

\bibliography{iterative}


\end{document}